\documentclass[12pt]{article}
\usepackage{graphicx}
\usepackage{amssymb,amsmath,amsfonts,palatino,amsthm}
\usepackage{amssymb}
\usepackage{epstopdf}
\newcommand{\f}{\begin{equation}}
\newcommand{\ff}{\end{equation}}

\usepackage{color}

\begin{document}
\title{The universe as a process of unique events}
\author{Marina Cort\^{e}s${}^{1,2,3}$ and Lee Smolin${}^{1}$
\\
\\
Perimeter Institute for Theoretical Physics${}^{1}$\\
31 Caroline Street North, Waterloo, Ontario N2J 2Y5, Canada
\\
\\
Institute for Astronomy, University of Edinburgh ${}^{2}$\\
Blackford Hill, Edinburgh EH9 3HJ, United Kingdom
\\
\\
Centro de Astronomia e Astrof\'{i}sica da Universidade de Lisboa${}^{3}$\\
Faculdade de Ci\^encias, Edif\'{i}cio C8, Campo Grande, 1769-016 Lisboa, Portugal}
\date{\today}

\maketitle

\begin{abstract}
We describe a new class of models of quantum space-time based on energetic causal sets and show that
under natural conditions space-time emerges from them.  These are causal sets whose causal links are labelled
by energy and momentum and conservation laws are applied at events.  The models are motivated by principles
we propose govern microscopic physics which posit a fundamental irreversibility of time.  One consequence 
is that each event in the history of the universe has a distinct causal relationship to the rest; this requires a novel form of
dynamics which an be applied to uniquely distinctive  events.  

We hence introduce a new kind of deterministic dynamics for
a causal set in which new events are generated from pairs of progenitor events by a rule which is based on  extremizing the
distinctions between causal past sets of events.  This dynamics is asymmetric in time, but we find evidence from numerical simulations  
of a $1+1$ dimensional model, 
that an effective dynamics emerges which restores approximate time reversal symmetry. 
Energetic causal set models differ from other spacetime-free causal set approaches, e.g.~Ref.~\cite{fotini} proposed causal sets based on quantum information processing systems, and Ref.~\cite{cohl} proposed causal sets constructed out of standard model particles.

Finally we also present a natural twistorial representation of energetic causal sets.

\end{abstract}

\newpage

\tableofcontents

\section{Introduction}

{\it "Time is an illusion. It emerges as we coarse grain, zooming out from timeless fundamental physics.  It's not a fundamental quantity of nature. In searches for quantum gravity we are to seek an underlying Planck length description, expressed in the form of timeless equations, where time is not to play any role. \\

When time emerges it is to parameterize equations that are symmetric under time reversal.  The evident time-asymmetry of  nature is then held to be an accident, due to an improbable choice of initial conditions." }\\

Such is the view that informs much work in quantum gravity and cosmology\footnote{We'll be referring throughout to ``cosmology'' as short for foundational cosmology, that is the discipline of the universe as a whole, not the traditional field of cosmology that deals with (large) universe subsytems.}. It is a view that twice diminishes our basic experience of the world as an unfolding series of moments, to the realm of accident and illusion: (1) because the asymmetry of time is held to be an accident, and (2) because time itself is held to be emergent. \\

In this paper we'll propose the diametrically opposite view. We develop the hypothesis that time is both fundamental and  irreversible, as opposed to reversible and emergent.   We'll argue that the irreversible passage of time must be incorporated in fundamental physics to enable progress in our current understanding.  The true laws of physics may evolve in time and depend on a distinction between the past, present and future, a distinction which is absent in the standard block universe perspective.  \\

The present instant is a primitive and part of fundamental processes, and the laws of physics may refer to it preferentially.  There is a process continually acting in the present bringing into existence the next moment. The present may code aspects of past states but the past is no longer accessible, as could be argued to be the case in the block universe picture. Along the same lines the future has yet to happen, and aspects of it may even be open, i.e. not computable from a complete knowledge of the present.   \\

We'll argue that, in contrast to the time reversal symmetry which is standard practice, time is fundamentally asymmetric and irreversible. The future is different from the past: the process by which the present becomes the past and gives rise to the future cannot be inverted to allow the perfect reconstruction of the past.

Based on this view we propose four principles.  The first wo concern the nature of time.

\begin{itemize}




\item{}{\bf Principle A}\\ Time is a fundamental quantity; the agency of time is the most elementary process
 in physics, by which new events are created out
of present events.   Causality results directly from the irreversible agency of time.

\item{}{\bf Principle B}\\ Time has a fundamental directionality. The future develops out of the present constantly; there are no causal loops and no regions or phenomena where time "evolves backwards."  This implies that the fundamental laws that evolve the future from the past are {\it irreversible} in the sense that they have no inverse by which the past state can be reconstructed from the present state.


\end{itemize}

A second pair frame the way that the dynamics of the world may be expressed.

\begin{itemize}

\item{}{\bf Principle C}\\ We choose a relational point of view, according to which the space-time properties of an object or event arise from its relationship with other objects or events.  All space-time properties have a dynamical origin.

\item{}{\bf Principle D}\\ Energy is fundamental.  Energy and momentum are not emergent from space-time, rather the opposite is the case, space-time is emergent from a more fundamental causal and dynamical regime in which energy and momentum are primitives.

\end{itemize}

We chose to model these assumptions within a discrete framework.   This means that we envision the history of the universe as a set of events, endowed with both a causal structure and intrinsic energy-momentum variables.

Before presenting this model, we note that the assumption that events are described relationally has a strong consequence, which is the

\begin{itemize}


\item{}{\bf Uniqueness of events} The relational properties of each event in cosmological evolution make it \textit{unique} and distinguishable from all others.

\end{itemize}

This is a consequence of the demand that each event be distinguishable by its relational properties.  Within a discrete causal structure an event can only be distinguished by its causal past.  Furthermore the events that make up the causal past of a given event cannot have any absolute labels, for those are only labeled by their causal pasts.  But those are part of the causal past of the given event.  This implies that any two events must have non-isomorphic causal pasts in that there is no map that takes one to the other preserving their causal pasts.  This can be considered to be a consequence of Leibniz's principle of the identity of the indiscernible. \\

%


Uniqueness needn't contradict our standard scientific method which we base on statistical inference from repeated experiments. Repeated systems do occur-to a sufficient degree of approximation- if we're considering subsets of the whole universe. Whenever we apply a boundary to define the system under study in the laboratory, we are severing the relations it has with the remainder of the system. Truncating these connections makes it possible for the subsystem to appear approximately similar to other subsets as well as to itself when subject to repetitions in time of the initial conditions.  
Similarity is local,  and a result of truncation, and repetition is never exact when the whole system is considered. \\

Uniqueness of events is a strong requirement as it demands each event has a  causal past that is complex enough to be district from the causal pasts of all the other events. As a result, fundamental laws acting on unique events need to take their complexity into consideration. This would seem to imply large informational inputs and outputs, which appears to contradict the idea that elementary events should be simple. We will show below that this query has a simple answer, which is that a space-time must emerge within which the network of complex historical relations  embeds. When an embedding of the history of events in a space-time exists one can use coordinates on the space-time to uniquely identify each event.   
 This is a highly non-trivial requirement, but in the next section we describe a model  in which it is satisfied.  \\


Having set out the physical picture which follows from our assertions, we present in Section~\ref{model_define} a simple model which is useful to study them. In this model a causal set\cite{cs} is generated by an event generator which acts according to a rule satisfying the principles we laid out.  The events are endowed with energy and momenta which are propagated from old to new events subject to conservation and energy-momentum relations.  We propose to call causal sets so endowed {\it energetic causal sets.}  We find some remarkable results:

\begin{enumerate}


\item{}Spacetime is not part of the fundamental description but emerges as an arena for a statistical description of the fundamental processes.

\item{}The emergent equations of motion, which describe the embedding of the causal processes in space-time, cannot always be solved consistently but when they can there is a classical limit which is a theory of interacting relativistic particles.

\item{}Numerical studies of the model in $1+1$ dimensions show that a reversible effective dynamics can emerge from irreversible evolution rules.  

\item{} In $3+1$ dimensions the model can be reformulated in the language of twistor theory.  

\end{enumerate}

In Section~\ref{model_define} we propose a model of causal time evolution and describe how space-time and the dynamics of relativistic particles are emergent from it.  Section~\ref{twistor} gives a reformulation of the model in the language of twistor theory, and in Section~\ref{1+1}  we consider the particular case of a $1+1$ dimensional model. We present numerical results from its study in Section~\ref{model_results}, that show the emergence of time reversible approximate laws from time irreversible evolution rules.  Our conclusions and prospects for further work are summarized in the last section.
In one companion paper \cite{motivation} we motivate and develop further each of the assertions here summarized, while in another \cite{halfquantum} we describe a quantum version of an energetic causal set and draw out some implications for quantum foundations.

The model we discuss in the next section is, so far as we know, new, but it relates in different ways to three earlier theories.  Causal sets have been studied intensively\cite{cs}, but usually they are conceived of having only  causal relations, from which all other properties including space-time geometry, momentum, etc. are hypothesized to be emergent.  Our energetic causal sets differ from them in that events and causal processes have intrinsic properties such as energy and momenta.  However, in two prior works, C. Furey proposed causal sets with intrinsic properties, in which the  causal links are conceived as constituted by elementary particles carrying various charges, from which space-time and fields are to be emergent\cite{cohl}.  Then, relative locality models\cite{rl} describe a world in which energy and momentum are prior to space-time; what we describe her evan be thought of as a kind of pre-geometry for relative locality.

\section{Energetic causal sets: A model of unique events} \label{model_define}

We now develop a model designed to investigate the principles we have just motivated. We should emphasize that we are not arguing this to be the unique model describing fundamental laws in a manner consistent with the principles we have proposed. Instead our aim is to study the nature of the dynamics emerging from the set of principles we proposed.

We model the history of a universe is described as a set of $N$ events, ${\cal E}_I$, $I=1, \dots, N$.  

Principle {\bf A} is incorporated here, in that each event is created as the result of a process acting on the prior set of events.  That process is the { \it activity of time.} In the model we call it
the {\it  events generator}.

The event generator must make two decisions each time it creates an event: 
First, which of the prior set of events are to be progenitors of the new event.  Second, how are the properties of the new event determined from the properties of its progenitors.

Each event is endowed with energy and momenta, which are held to be primitive properties.   These are conveyed to an event from its progenitors.
This realizes Principle {\bf D}.

Each event has several incoming momenta
and several outgoing momenta.  
$p_{aK}^I$ is the momenta incoming to event $I$ from event $K$ and $q_{aI}^L$ is the momenta outgoing from event $I$ to event $L$.
$a=0,\dots,3$ is an index in momentum space, which we assume has a minkowski metric, $\eta^{ab}$.

The momenta are subject to three sets of constraints.
\begin{enumerate}

\item{}Conservation laws:
\f
{\cal P}_a^I = \sum_K p_{a K}^I -  \sum_L q_{a I}^L  =0 
\ff
where the sum over $K$ is over all events $I$ is connected to in the past and the sum over $L$ is over all events $I$  is connected to in the future.
\item{} No redshifts
\f
{\cal R}_{aI}^K = p_{aI}^K - q_{aI}^K =0 
\ff
\item{}Energy momentum relations for massless photons
\f
{\cal C}^I_K = \frac{1}{2} \eta^{ab} p_{a K}^I p_{b K}^I  =  0 , \ \ \ \ \  \tilde{\cal C}^I_K = \frac{1}{2} \eta^{ab} q_{a K}^I q_{b K}^I  =  0 
\label{em}
\ff

\end{enumerate}

Together, the causal structure and the momenta are a complete description of the world in this model.  The dynamics
is completely given by the event generator which picks the progenitors of new events and these three sets  of constraints.

Finally, let us emphasize that the metric, $\eta^{ab}$ that occurs in (\ref{em}) is the metric on momentum space.  Fundamentally  there is no
space-time and hence no space-time geometry.  One could also take the momentum space geometry to be other than flat, in this case
one would have a model of relative locality\cite{rl}.

\subsection{Statistical description}

We can gain more insight into the physics  of this model by introducing a statistical formulation.   Since the dynamics which generates the causal set is deterministic, but non-local, we have to define carefully what ensemble the statistical averaging corresponds to.  We also have to respect the assumption that each fundamental event is unique when all the information about its causal past is known.  There is also a single, unique causal history for the whole universe, ${\cal C}^{universe}$.
A statistical ensemble can then only arise from the consideration of subsystems of intermediate scale, neither cosmological nor universal.
We propose then to introduce statistics by studying ensembles composed of subsystems which are characterized by incomplete information about their causal pasts.

Let us then consider a class of subsystems of the universe, which is defined by a fixed number of incoming and outgoing momenta, $n_{in}$ and $n_{out}$.    To further specify the ensemble, we can fix the values of the incoming and outgoing momenta, $ p_{a}^{in}, p_a^{out} $.  Within the single history of the entire universe, the subsystems which match these initial and final conditions  constitute\footnote{Note that energy-momentum conservation may impose conditions on the incoming and outgoing momenta.  If there are no redshifts, for example, the total incoming momenta must equal the total outgoing momenta.} an ensemble, ${\cal E}(n_{in},n_{out}; p_{a}^{in}, p_a^{out} )$.

Each element of the ensemble has a causal structure, $\cal C$, which is a sub-causet of the full causal set of the universe, ${\cal C}^{universe}$,  as well as particular values of energy-momenta propagating on its internal links.  

We consider an observable,  ${\cal O}[p_a^{IJ}, {\cal C}]$, which may be a function of the intermediate causal structure and momenta variables.
We wish to give an estimate for its average value  in the ensemble, ${\cal E}(n_{in},n_{out}; p_{a}^{in}, p_a^{out} )$.
In general this will depend on the detailed causal history of the whole universe.  This reflects the fact that the causal structure and intermediate momenta of a subsystem depend on the detailed causal past of that subsystem.  However we may introduce here an hypothesis that {\it the universe is generic on intermediate scales}.  This means that in the ensemble, ${\cal E}(n_{in},n_{out}; p_{a}^{in}, p_a^{out} )$, all causal structures and intermediate momenta consistent with the constraints occur with equal probability\footnote{Within a given causal history of the universe, the ensemble, ${\cal E}(n_{in},n_{out}; p_{a}^{in}, p_a^{out} )$, is real so these are relative frequency probabilities.}.  We call this the {\bf meso-generic principle.}

The key point is that  that knowledge of the incoming and outgoing momenta does not suffice to determine the causal structure of the intermediate scale subsystem.  This is because the causal structure in a region is determined by knowing the pasts of the incoming processes, not just their momenta.  The way that we have defined the ensemble, ${\cal E}(n_{in},n_{out}; p_{a}^{in}, p_a^{out} )$, we are ignorant of this information.  (Indeed were the pasts of the incoming processes completely specified we would not have an ensemble, but rather a single, unique process.)  Hence if we want to compute probabilities relevant for local observables with incomplete, local, information, we must average over the causal structures and internal momenta that could evolve the $n_{in}$ processes to the $n_{out}$ processes.  The new 
 {\bf meso-generic principle} expresses the idea that each possible choice of internal causal structure and momenta occurs equally often in the ensemble, i.e. the mesoscopic physics is maximally decoupled from the single, unique causal history of the whole universe.  Given this, we can define local expectation values
\f
< {\cal O}   >_{\cal E} (n_{in},n_{out}; p_{a}^{in}, p_a^{out} )=  \sum_{\cal C}   \int \Pi_{(IJ)} dp_a^{IJ} dq_a^{IJ} \delta ({\cal C}_a^{IJ} ) \delta ({\cal R}_I^J ) \Pi_{I }  \delta ({\cal P}_a^{I} )  {\cal O}[p_a^{IJ},  {\cal C}]  
\label{stat-local0}
\ff
\begin{eqnarray}
< {\cal O}   >_{\cal E} (n_{in},n_{out}; p_{a}^{in}, p_a^{out} )&=& \nonumber\\ && \sum_{\cal C}   \int \Pi_{(IJ)} dp_a^{IJ} dq_a^{IJ} \delta ({\cal C}_a^{IJ} ) \delta ({\cal R}_I^J ) \Pi_{I }  \delta ({\cal P}_a^{I} )  {\cal O}[p_a^{IJ},  {\cal C}]  \nonumber
\end{eqnarray}

To compute (\ref{stat-local0}),  we  introduce some lagrange multipliers to exponentiate the constraints.  
\f
< {\cal O}   >_{local} =  \sum_{\cal C} N [ {\cal C}] \int \Pi_{(IJ)} dp_a^{IJ}  dq_a^{IJ} d{\cal N}_I^J  d \tilde{\cal M}_I^J  \Pi_{I } dz^a_I  \ e^{\imath S^0}  \ 
{\cal O}[p_a^{IJ}, {\cal C}]  
\label{stat1}
\ff
where $S^0$ is a dimensionless  action
 \f
 S^0= \sum_I z^a_I {\cal P}_a^I   +\sum_{(I,K)} (   x^{a I}_K {\cal R}_{aI}^K + {\cal N}^K_I {\cal C}^I_K -  \tilde{\cal N}^K_I  \tilde {\cal C}^I_K )
 \label{S0}
 \ff
where the sum over $(I,K)$ is over all connected pairs of events.  

There are three kinds of lagrange multipliers, $z^a_I$ associated with each event, $x^{a K}_I$ and 
 the ${\cal N}^I_K$ and  $\tilde{\cal N}^I_K$ are associated with each connected pair of events. 

It is natural to give the  $z^a_I$ and the  $x^{a K}_I$ dimensions of inverse momenta.  

Note that, in the absence of information that could determine a non-constant weight $ N [ {\cal C}] $,  the action arises entirely from the representation of the constraints by integration over lagrange multipliers.



\subsection{The emergence of space-time}

Let us now consider that the products $x \cdot p$ and $z \cdot p$ in the action (\ref{S0}) are large compared to unity so that we can evaluate the integrals in
(\ref{stat1}) in the stationary phase approximation.  We then seek the critical points of $S^0$.  We will see that this leads to the emergence of space-time.  

The variation of the action by the lagrange multipliers gives the constraints.  But we have new equations satisfied by the lagrange multipliers coming from the variation of the action by the momenta.  
\f
\frac{\delta S^0}{\delta p_{a K}^I }= z^a_I + x^{a K}_I + {\cal N} p^{a I}_K =0 
\label{z+x}
\ff
\f
\frac{\delta S^0}{\delta q_{a  I }^K } = -z^a_K - x^{aK}_I -  \tilde{\cal N} q^{a K}_I =0 
\label{-z-x}
\ff

Adding these two equations and using $ {\cal R}_I^K =0 $ we find
\f
 z^a_I -  z^a_K =  p^{a I}_K (\tilde{\cal N}_I^K  - {\cal N}_I^K )
 \label{emergence}
\ff
This has a simple physical interpretation.  The lagrange multiplier $z^a_I$ can now be interpreted as the  space-time coordinate of the
event $I$.  $ z^a_I -  z^a_K $ is then a space-time interval between event $K$ and event $I$.  It is a light-like interval proportional to
the momentum $p^{a I}_K$ connecting $K$ to $I$.  The constant of proportionality involves the lagrange multipliers  $\tilde{\cal N} - {\cal N} $ which is consistent with the fact that the affine parameter along a null ray is arbitrary.  

We can also take the difference between (\ref{z+x}) and (\ref{-z-x}) to find equations that determine the lagrange multipliers $x^{a K}_I$,
\f
x^{a K}_I= \frac{1}{2} (   z^a_I + z^a_K   + p^{a I}_K (\tilde{\cal N}_I^K  + {\cal N}_I^K ) ). 
\ff
These may be given a physical interpretation, identifying the causal process with an event displaced from the average of the embedding coordinates of the two events it connects \footnote{This is somewhat reminiscent of the equations of relative locality\cite{rl}.}, but this plays no role in what follows.

We may note that equations (\ref{emergence}) will not always have simultaneous solutions.  There is one equation to be solved for every causal link, but only one $z_I^a$ for each event.  In the cases where the equations can always be solved it means that there are a consistent choice of embeddings
$z_I^a$ of the events in a flat space-time such that the causal links are represented by null intervals proportional to the energy-momentum they carry.
In these cases, 
we can say that {\it space-time has emerged,}  as there was no space-time and no locating the events in space-time in the original description.  It is interesting to note that the emergent space-time inherits the metric $\eta^{ab}$ from momentum space. 

The lagrange multipliers $z^a_I$ start off as just arbitrary variables with no physical meaning other than reinforcing the constraint.  By virtue of the variational principle which says the action is an extremum under all variations, the $z^a_I$ become coordinates embedding the events in
Minkowski spacetime-where the metric of space-time comes from energy-momentum space.

We should caution that the analysis we are giving here is crude, as we are just relying on the stationary phase approximation, without taking into account measure factors that could arise either from a strongly varying $ N [ {\cal C}] $ or from the integral over momenta.  But within this simple approximation we can draw a striking conclusion, which is that the causal histories that dominate the ensemble are those whose causal structures and momenta are chosen so that the 
$z^a$ do embed in a flat spacetime.

Note, finally, that these coordinates $z^a_i$ have units of inverse momenta, which is not conventional for embedding coordinates in space-time.  If we wanted to introduce conventional units of length we could rescale the $z^a_i$ by a constant with dimensions of action, $\hbar$.  
The introduction of $\hbar$ is purely conventional and arises only because we wish to give these lagrange multipliers units of length rather
than inverse momenta.  There are no quantum amplitudes and probabilities are conventional classical 
probability weights\footnote{For the corresponding quantum formulation see \cite{halfquantum}.}.

\subsection{The solution to the question of dynamics for unique events}

The emergence of the $z_I^a$'s as coordinates of the embeddings of the events in a space-time neatly solves the puzzle of how to give dynamics to unique events.  As we posed the question in the introduction, the question was how to conceive of a law or a rule for generating unique events when that rule had to be simple, but what distinguishes the events in a big universe are the intricacies of their histories. The answer is to invent space-time so that each event has a unique embedding in that spacetime.  

The $z^a_I$'s transform under the symmetries of space-time but invariants of intervals between them are relational observables.  They summarize complex combinations of the past of the events, taking into account the flow of energy-momenta as well as the event generator.  So the answer to the puzzle is that the dynamics can depend on the $z^a_I$'s.

\subsection{The emergence of relativistic particle dynamics from chain of events}

We can consider the example of a chain of events ${\cal E}_I$, $I=1,É N$, which each have a single incoming and single outgoing momenta, denoted simply by $p_a^I $ and $q_{a I}$.  
Alternatively, we can regard these as a chain of interactions at which all but one incoming and one outgoing momenta are negligible.

We can solve the $\cal R$ constraints to find 
\f
p_a^I = q_a^I 
\ff
We then have for the $z^a_I$,
\f
 z^a_{I+1} -  z^a_I =  p^{a I} {\cal M}_I
\ff
where 
\f
{\cal M}_I =  \tilde{\cal N}_I^{I-1} - {\cal N}^{I+1}_{I}
\ff
Now we can imagine that ${\cal M}_I$ are small so the adjacent events in the chain are close to each other. So we can expand
\f
z^a_{I+1} = z^a_{I} + \dot{z}^a (t) \Delta t
\ff
where $\Delta t$ is a small interval.  We then have
\f
\dot{z}^a (t) = \frac{{\cal M}_I}{\Delta t} p_I^a = n p_I^a
\ff
where $n= \frac{{\cal M}_I}{\Delta t} $ remains finite as both $\Delta t $ and  ${\cal M}_I$ are taken to zero.

Note that we have used the $\cal R$ constraints and so eliminated their lagrange multipliers $x^a_I$ but we have not yet
used the $\cal P$ constraints or the $\cal C$ constraints.   The action is then
\f
S = \sum_I p_a^I (z_{I+1}^a - z^a_I ) - \frac{1}{2}  {\cal M}_I p_I^2 
\ff
If we take the limit of $\Delta t \rightarrow 0$ so that $\sum_I \Delta t \rightarrow \int dt$ so that the chain goes over in the limit to a curve.
In this limit the $p_a^I$ can be replaced by the continuous functions $p_a(t)$.  We will also replace the
discrete $z_I^a$ by continuous variables $x^a (s)$ so that
the action becomes
\f
S^{free} = \int dt \left ( p_a (t) \dot{x}^a (t)  - \frac{1}{2}  n (t)  p (t)^2 \right )
\ff
which is the action for a free relativistic particle.  

Note that this form is invariant under reparameterizations 
\f
t \rightarrow t^\prime =  f(t) , \ \ \ \ \  dt^\prime = \dot{f} dt ,\ \ \ \ n (t) \rightarrow n^\prime (t^\prime ) = \frac{n(t)}{\dot{f}}
\ff
$n(t)$ is a lagrange multiplier that gives the energy-momentum relation as a constraint, $p^2=0$

By an integration by parts we have (neglecting the boundary terms)
\f
S^{free}= \int dt \left ( - x^a (t) \dot{p}_a (t)\  - \frac{1}{2}  n(t)  p (t)^2 \right )
\ff
The $x^a (t) $ are then lagrange multipliers that enforce the equation of motion for a free particle
\f
\dot{p}_a (t)\ =0
\ff
while the variation by $p_a (t)$ gives the relation between the velocity and momenta
\f
\dot{x}^a = n (t) p^a
\ff
where the lagrange multiplier is arbitrary reflecting the reparameterization invariance.

Next we consider a network of long chains connected together by intersections of three or more chains.  We continue to label these 
intersections by $I$ and the chains that connect them by $(I,J)$.  Taking again the limit of $S^0$ for each chain we
have
\f
S^0 \rightarrow S^{rel} = \sum_{(I,J)} S^{free}_{(I,J)} + \sum_I z^a_I {\cal P}^I_a  ,
\label{Srel}
\ff
where the conservation law ${\cal P}^I_a$ is a function of the momenta at the endpoints of the paths that meet at the intersection
point $I$.

$S^{rel}$ given by (\ref{Srel})  is the action for a process in which a set of free relativistic particles interact at intersections where the conservation laws
${\cal P}^I_a$ are satisfied.  Note that the equations of motion for the $p_a(s)$ at the end points enforce the locality of the interactions by
equating $z^a_I$,  the coordinate of the $I$'th intersection with the coordinate of the endpoint of the path that meet there,
$x^a (s=1) $ or $x^a (s=0) $.

Before closing this section, we should point out that our discussion of the emergence of space-time and particle trajectories has relied entirely on the principle of stationary phase.  We have not so far attempted to estimate the measure factors that govern the relative importance of different critical points or can even lead to the dominance of the path integral by non-critical histories.  This is a good problem for future work.

\section{Twistor formulation}\label{twistor}

We now describe an alternative formulation of the $3+1$ dimensional model related to twistors.  We can solve the energy-momentum
constraints for massless particles (\ref{em}) in terms of two component spinors. We represent null $p_a^{IJ}$ by a two component spinor $\pi^{IJ}_A$ by
the correspondence
\f
p_a^{IJ} \leftrightarrow \bar{\pi}_{A'}^{IJ} \pi_A^{IJ} 
\ff
ie $p_a = \sigma_a^{A'A} \bar{\pi}_{A'} \pi_A $ for $3+1$ dimensional Pauli matrices, $ \sigma_a^{A'A}  $.

$q_a^{IJ}$  are similarly represented by spinors $\chi_A^{IJ}$.  
\f
q_a \leftrightarrow \bar{\chi}_{A'}^{IJ} \chi_A^{IJ} 
\ff
The  redshift constraints are now
\f
{\cal R}_{AI}^K = \pi_{AI}^K - \chi_{A I}^K =0 
\ff
The conservation law at each event is
\f
{\cal P}_{AA'}^I = \sum_K \pi_{A K}^I  \bar{ \pi}_{A' K}^I -  \sum_L \chi_{A I}^L  \bar{\chi}_{A' I}^L =0 
\ff

We again form an action to express the constraints.  The energy-momentum relation constraints are not present because they are solved for.
 \f
 S^{twistor}= \sum_I z^{AA'}_I {\cal P}_{AA'}^I   +\sum_{(I,K)}    \lambda^{A I}_K {\cal R}_{A I}^K  
 \ff
The variation of the action by ${\pi}_{A I}^K$ yields the twistor equation
\f
 \lambda^{A I}_K = z^{AA'}_I \bar{\pi}_{A' K}^I
\ff

The interpretation of this is the following\cite{twistors}.
The $\pi_A$ denote null directions, while the pair $Z = (\bar{\pi}_{A'} , \lambda_A )$ denote null lines in Minkowski spacetime. 
A $z^{AA'}$ satisfying the twistor equation is an event on that null line.  Now there is one pair $Z_I^J$ for every causally connected pair
of events and hence a null line for each pair.  If an event has $M$ null lines through it its corresponding $z^{AA'}$ satisfies $M$
twistor equations, this means that that event lies at the intersection of those null lines.

So we elegantly reconstruct the embeddings of the causal processes in Minkowski space-time.

\section{$1+1$ dimensional model}\label{1+1}

We now focus on a model in $1+1$ dimensions so that space-time indices $a,b,c, \ldots$ take values $0,1$. We first give a detailed description of the model and then introduce
an algorithm by means of which it may be simulated.  The results from a set of numerical experiments will be described in the next
section.

We should emphasize that we choose the set up of this simple model so that the equations (\ref{emergence}), that define the embedding of events in a $1+1$ dimensional space with Minkowski metric, are always satisfied.  This model cannot thus be used to illustrate the issues that arise in the general case when the emergence of space-time may be less automatic.  This does make the model useful to investigate the issue of how an effectively reversible dynamics may emerge from  a fundamental irreversible system, and that is its main purpose.

\subsection{Description of the model}

We will consider events that have two incoming photons and
two  outgoing photons, in which one of each pair is left-moving, the other is right moving.  On shell momenta are then given by a value of
energy, so $p_a^R = (E, E)$ while $p_a^L = (E, -E)$.  

The energy incoming into event $I$ will be denoted $E^{L,R}_I.$  The corresponding two-momenta are $p^{L,R}_{Ia}.$
The energy out going from event $I$ will be denoted $Q^{L,R}_I$.  The corresponding two-momenta are $q^{L,R}_{Ia}.$

Thus, we have a conservation law
\f
P^I_a = p^L_{I a} +  p^R_{I a} = q^L_{I a} +  q^R_{I a}= (E^{L}_I+  E^{R}_I,  E^{R}_I - E^{L}_I ) = (Q^{L}_I+  Q^{R}_I,  Q^{R}_I - Q^{L}_I ) 
\ff
Below we will want the emergent space-time to be spatially compact.  This requires 
 the spatial momenta come in integral units of a fundamental momenta,  $\epsilon$, so
\f
E^{L,R} = k^{L,R} \epsilon, \  \  \  \  Q^{L,R} = q^{L,R} \epsilon,
\ff
where $k^{L,R}$ and $q^{L,R}$ are non-negative integers.

The constraints can be expressed as conditions on the integers $k^{L,R}$.
\f
\tilde{\cal P}_a = \frac{{\cal P}_a }{\epsilon} = (k_L + k_R -q_L -q_R , k_L - k_R -q_L +q_R ) = 0
\ff

The path integral now takes the form
\f
< {\cal O}   >_{local} =  \sum_{\cal C}   N [ {\cal C}]  \int \Pi_{(IJ)} dp_a^{IJ} dq_a^{IJ} \delta ({\cal C}_a^{IJ} ) \delta ({\cal R}_I^J ) \Pi_{I }      \delta_{\tilde{\cal P}_0} \delta_{\tilde{\cal P}_1} 
\delta ({\cal P}_a^{I} )  
{\cal O}[p_a^{IJ},  {\cal C}]  
\label{stat-local1+1}
\ff
where the delta functions for energy-momentum conservation at each event have been replaced by Kroneker delta's.

We exponentiate the constraints with a compact integration over the Lagrange multiplier $z^a$ imposed in the spatial direction,
\f
\delta_{\tilde{\cal P}_0} \delta_{\tilde{\cal P}_1} = \int dt \oint_0^L dx e^{\imath \epsilon z^a \tilde{\cal P}_a }
\ff
where we have imposed
\f
z^a = (t,x)  \sim (t,x+L)
\ff
where 
\f
L  = \frac{2 \pi}{\epsilon}.
\ff

The spatial compactness is also helpful to get an interesting model, without it the worldliness run away from each other as nothing happens after a finite number of interactions.

The space-time coordinates of the events will be related by the equation of motion coming again from the stationary phase approximation 
\f
 z^a_I  =  z^a_K  + \eta^{ab}   p^{ I}_{ b K}  \tilde{\cal M}_I^K  
\ff

It is very convenient to go to null coordinates
\f
z_\pm = ( t \pm x )
\ff
so we write $z^a = (z_+, z_- )$.  Left moving null curves are labeled by $z_+ = constant$, right moving null
curves by $z_- = constant$. 
The metric is
\f
ds^2 =  dz^+ dz^- 
\label{metric}
\ff

 The periodic boundary conditions are now
\f
z^a = (z_+ , z_- ) \sim (  z_+ +nL  , z_-   - n L ),
\ff
where $n$ is an integer.
The first continuation of a left moving null trajectory with $z_+ = a$ will be denoted $z_+ = \tilde{a}$. The next by
 $z_+ = \tilde{a}^{(2)}$ and so on.

The equation of motion is then, for $p^a=(p_+, p_-)$
\f
p_- = \frac{2}{\cal M} ( z_1^+ - z_2^+ ) , \ \ \ \ \ \ \  p_+ = \frac{2}{\cal M} ( z_1^- - z_2^+) 
\ff
For a null line one of these vanishes, because either $z_+$ or $z_-$ is conserved along it.

The conservation laws are very simple in null coordinates
because, from the equations of motion,  $p_a^R = (0, k_R \epsilon ) $ for right movers, and $p_a^L = (k_L \epsilon, 0 ) $ for left movers.
Therefor momenta just continue through each event:
\f
q^I_L = k^I_L, \ \ \ \ \   q^I_R = k^I_R, 
\label{em1+1}
\ff

The equations of motion then just determine the lagrange multipliers $\cal M$.

Given any two events $z^a_1 = (z_1^+,z_1^-)$ and $z^a_2 = (z_2^+,z_2^-)$ with $z_1^+<z_2^+$ there will be an event where the right going null
ray from $z_1$ intersects with the left going null ray from $z_2$. This takes place at $z_3^a = (z^+_2,z^-_1)$. Correspondingly the left going ray from $z_1$ intersects the right going ray from $z_2$ at  $z_4^a=(z^+_1,z^-_2)$. Generically photons moving away from each other (towards the boundaries) inherit the outward coordinates of the parent pair of events, and photons moving towards each other inherit the inward coordinates of the parent pair. \\
This yields the generic rule for generation of new events,

\begin{equation}
(z_{\rm new}^+,z_{\rm new}^-)=(z_{(1,2)}^++n\,{\rm L},\,z_{(2,1)}^--m\,{\rm L})\,,
\label{interaction}
\end{equation}

where subscripts refer to the pair generating events with $z^+_1<z^+_2$ and the order in the subscript $z^{\pm}_{(1,2)} $ refer to outward or inward moving photons, respectively. $n$ and $m$ need not necessarily be identical, two photons can undergo a different number of cycles before interaction. In terms of space time coordinates this means,

\begin{eqnarray}
\begin{cases}
 t_{\rm new}=\frac{1}{2}\left(t_1+t_2\pm x_1\mp x_2+ (n-m) L \right) \,,\\
 x_{\rm new}=\frac{1}{2}\left( \pm t_1\mp t_2+ x_1+ x_2+ (n+m) L \right) 
 \end{cases}
 \label{mixup}
\end{eqnarray}

where $x_1<x_2$ and upper and lower $\pm,\mp$ refer to outward and inward moving photons respectively.

For every pair of events, and due to periodicity, there are an infinite number of intersections of the continuation of their null lines. The full set of possible intersections is specified by $(n,m)$ given by
 
\begin{equation}
(n,m)=\{(n,n), (n,n+1), (n+1,n)\}\,,\label{nm}
\end{equation}
the two latter cases corresponding to the right or left moving photon crossing the boundary first. So photons in different windings can interact but the difference in windings of $L$ is limited.  \\

 
If we always pick the closest intersection of null rays as the next event, that is if $m=n=0$, there is no possibility of photons interacting in future cycles, there will be an interaction every time two rays cross. Since we are in $1+1$ dimensions this means the system will reduce to a simple model of photons oscillating back and forward, where each photon interacts alternately only with the two photons to its immediate left and right. So here we will introduce an amount of random input in the model by randomly taking $n=\{0,1\}$ in the generation of new events.

For constructing the model we will select which pairs of events interact according to a rule which takes into account some measure of
the differences in the past of each pair of events. The pasts will be described by positive numbers ${\cal D}_{IJ}$ which
measure how different are the pasts of events $I$ and$J$. For example if the past is a graph,
the difference between two pasts could be the norm of the differences of their adjacency matrixes.\\

However even a moderately complex measure of each pasts quickly becomes computationally unfeasible
 because the number of differences of pasts to compute in the generation of each new increases exponentially with the number of intervening events. Since our sole requirement for each past is that it be unique, the need for complexity can be alleviated by taking a simple measure of each as the average of the space-intervals of events in that past. This we know to be unique since we have continuum space-time available. So the emergence of space-time from the interaction momenta is at the same time ensuring uniqueness of each past.
Past will then be stored as 
\begin{equation}
{\rm past}_I^2=\frac{1}{N}\displaystyle\sum\limits_{J}(-t_J^2+x_J^2)\,,
\label{past}
\end{equation}
where $x_J$ is valued in compact space and $N$ is the number of events in the past of event $I$ (events in the sum in $J$). Each event is distinguished by its past, as well as its position and time labels, and is stored in the past of the parent event that contributed the left incoming ray \footnote{We have studied models with the opposite continuation rule - retracing the right incoming ray - and tested that in none of the relevant cases any asymmetry arises between left and right evolution.}. Hence all events belong to one and only one past, and the intersection of all pasts is null. The distance between pasts will be, 
\begin{equation}
{\cal D}_{IJ}=|{\rm past}_I^2-{\rm past}_{J}^2|\,.
\label{distance}
\end{equation}

We're using the norm and not square root because it considerably reduces computational time, since the number of square roots being evaluated is order $10^7$. This is of no effect since we're concerned with differences between pairs and not the absolute value.
Given an expression for the distance between each pair of pasts, we can now determine how to rule their interaction. We will mostly be selecting pairs which satisfy

\begin{equation}
{\rm Interacting\, Pair} = {\rm Min}\{{\cal D}_{IJ}\}
\label{min}
\end{equation}
though in Section~\ref{max} we'll experiment with other forms for the selection rules. 

A last observation, before we go on to discuss the implementation of the model, is that the model as so far described has a conformal symmetry.   This is because the rule for generating events just makes use of the intersections of null curves in the emergent spacetime.  Thus, if we multiply the metric by a function $e^{ 2 \Omega (z^+ , z^- )}$ so that
(\ref{metric}) is replaced by,
\f
ds^2 = e^{ 2 \Omega (z^+ , z^- )} dz^+ dz^- 
\label{metric2}
\ff
then, sop long as $\oint dx  e^{ 2 \Omega (z^+ , z^- )} =1$ so that the overall length $L$ is preserved, 
nothing in the evolution rules or their consequences would change.

\subsection{The algorithm}\label{algorithm}

The dynamics of the model, i.e. the generation of new events, which we identify with the generation of time instants, are then reproduced in the following procedure:

\begin{itemize}

\item{}{\bf STEP 0}:  Initialize.  

Choose the total number of events to be generated events, $N_{events}$.
Pick $N_{past}$ initial events, at random spacial coordinates at global time $t = 0$, and pick $P_a^I = (p^I _L, p^I_R  )$ for $I=1,ÉN$. $N_{past}$ also corresponds to the total number of intervening pasts we'll keep track of. 

The pasts of the initial events are simply their current space-time position, given by Eq.~(\ref{past}).

\item{}{\bf STEP 1}: Create a new event:

Pick the open pair with the lowest value of ${\cal D}_{IJ}$ and construct the first event to the future of $I$ and $J$\footnote{This kind of extremal dynamics in which the next move involves a a choice of a subset that extremizes some observable is reminiscent of models of self-organized criticality like the Bak-Snepen model\cite{BS}.}.

An event is open, and thus candidate to interaction, if there aren't yet two null lines coming out of it to the future. 
The only pairs available for interaction are those that still have an opposite free pair of photons. Two events, both having only one free right moving momentum, cannot interact. 
If both events have both null rays unused, select randomly a pair of momenta for interaction.  
Randomly select $n$,  the number of cycles before interaction, and determine $m$ by Eq.~(\ref{nm}).

Create the two null lines that connect $I$ and $J$ each to the new event, hence determine the coordinates of the new event.  The momenta of these photons
are computed trivially by continuation, according to the conservation rule Eq.~(\ref{em1+1}).

\item{\bf STEP 2}: Compute the past of the new event and store it in the past of the parent that contributed the left incoming ray.

\item{\bf ITERATE} go back to STEP 1. Repeat $N_{events}$ times.

\end{itemize}

Note that the initial $N_{past}$ events fully specify the coordinates of all possible future events, and that all values of future momenta are specified by the conservation law, so the dynamics is deterministic, up to factors of the number of cycles a momentum executes before interaction, $n$ \footnote{If we want fully deterministic dynamics we can pick $n$ dependent on either 1) the difference of pasts of each pair of events; 2) the lagrange multiplier for each momentum.}.

Note also that because of the conformal symmetry, the $x$ coordinates of the initial events serve only to specify their order around the circle, beyond that their values don't matter.  

This model is very simple, the momenta play a small role, which is just to propagate themselves according to the conservation rules.   But it can be used to study
the question of how time reversible approximate laws might emerge out of time irreversible fundamental laws, and hence address our central assumption, principle {\bf A}.
The evolution rule appears to be very time asymmetric and in Section~\ref{max} we will play with different rules and different
amounts of time asymmetry-measured by how much information about the past they use or forget.  Then we can
see if averaged quantities come to any sort of time independence or equilibrium.  This and other questions will be examined in the next section where we present the simulation results.

\section{Simulations of the $1+1$ dimensional model}\label{model_results}

We studied the $1+1$ dimensional model we have just described in Section~\ref{algorithm} by numerical simulations.  We did many runs, varying the event generation rule,
the initial system size (the number of initial events which is the same as the number of distinct past sets) and the length of the run.  We start the system off with a set of initial events with random spatial positions at an initial time.

We first describe the results, then the details of the simulations

\subsection{Description of Results}

We saw evidence for a simple characterization of the time evolution in which 
the systems pass in time through two distinct phases.  The systems begin in a disordered phase, followed by an ordered phase we call the locked in phase.   In the disordered phase, the time asymmetry of the event generation rule manifests itself in a visible time asymmetry of the pattern of space-time positions of the events.  In this initial phase, the events form a roughly random pattern in space-time, characterized by a large variety of spatial positions, within an overall envelope that is asymmetric under time reversal.   In the locked in phase an approximate time symmetry emerges. This second phase is dominated by persistent repeating patterns we can call quasiparticles.     

We observed this two phase structure in runs with several distinct event generation rules.

Each run began with a period of seemingly chaotic behaviour, manifested by a  disordered ensemble of events in which the time asymmetry of the algorithm is evident.  In this disordered phase all the past-sets intervene with similar weights in the generation of new events. Variety is maximal.  Any two past-sets interact only very briefly -- two times -- just enough for both pairs of photons of the parents to be used and discarded, and interaction moves on to another pair.

\begin{figure}[t!]
\centering
$\begin{array}{cc}
\includegraphics[width= 0.5\textwidth]{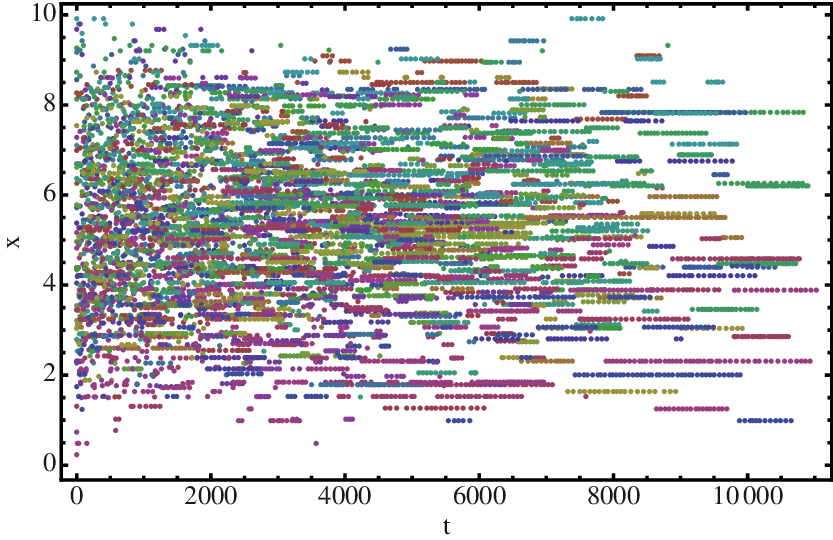}&
\includegraphics[width= 0.5\textwidth]{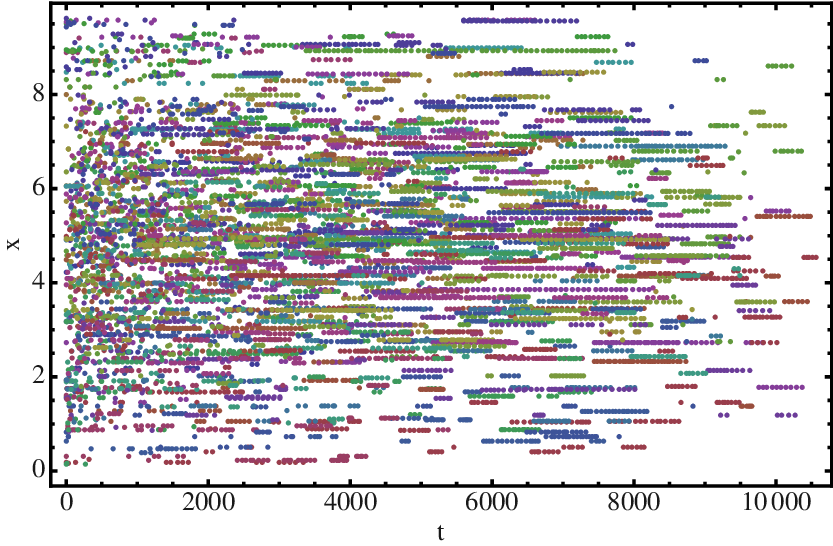}\\
\end{array}$
\caption{1+1 dimensional unique events model described in Sections~\ref{1+1}--\ref{model_results}.
Interaction of similar pasts, given by Eq.~(\ref{min}) between 40 past-sets and during $10^4$ events. We show two examples with the same parameters to convey variation between runs.
Each simulation begins with a period of disordered behaviour, in which the time asymmetry of the algorithm is evident. As time progresses, past-sets interlock briefly, and glimpses of stable trajectories, the quasi-particles emerge. Quasi-particles move in straight lines, and as such have dynamics which appears unchanged under time reversals. \textit{We say that time symmetric dynamics emerged from time asymmetric rules.}}
\label{fig1} 
\end{figure}

This behaviour is illustrated in Fig.~\ref{fig1}. We start with 40 past-sets and generate each run for $10^4$ events. We will show throughout two different runs with the same parameters to convey variation between different realizations. Each dot represents an event and its color indicates which past-set it is stored in. Within each plot different past-sets are distinguished by color, 
We choose to plot the process in space-time coordinates as opposed to null coordinates $z^{\pm}$ since null coordinates remain constant throughout evolution (up to cycles of $L$). 

In Fig.~\ref{fig1} we see a very different behaviour emerging after a sufficient number of events has been generated. As time progresses, we gradually begin to see two past-sets interlocking, ever so briefly for a short succession of events, and we start to see momentarily glimpses of stable trajectories in the plotted evolution. The interlocking between pairs of past-sets lasts longer each time it occurs.
These give rise to recognizable trajectories of quasi-particles.

Towards the end only a handful of past-sets remain, giving rise to stable space-time trajectories of emergent quasi-particles. These trajectories, can be described as the propagation of quasi particles consisting of long chains of events with one or more common past-sets. Quasi-particles move in straight lines in space-time, and as such have a dynamics which appears to be unchanged under reversal of the time coordinate.  We call this the \textit{lock-in phase}.

\begin{figure}[t!]
\centering
$\begin{array}{cc}
\includegraphics[width= 0.5\textwidth]{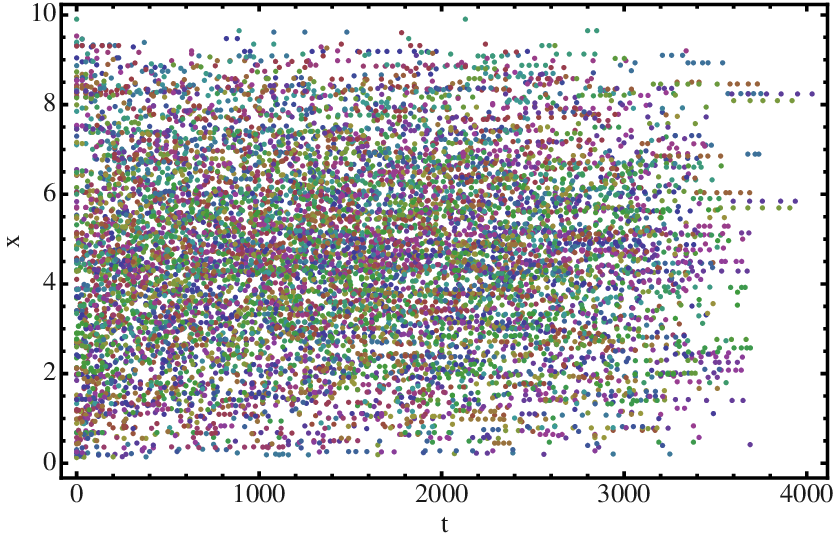}&
\includegraphics[width= 0.5\textwidth]{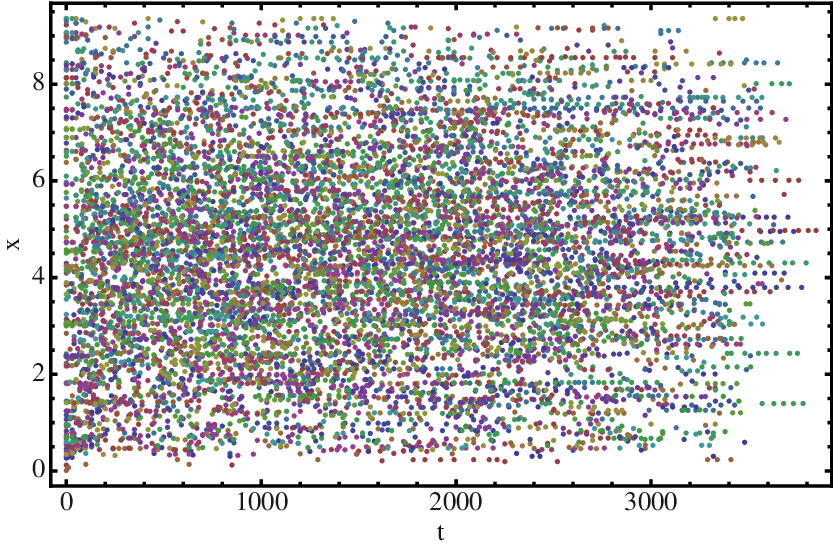}\\
\end{array}$
\caption{Interaction of similar pasts: 100 past-sets interacting for $10^4$ 
events. Same parameters as Fig.~\ref{fig1} but increased number of past-sets. Two examples of the same parameters show variation within different runs. The emergence of quasi-particles takes longer within the same number of events but still occurs.}
\label{fig2} 
\end{figure}

\begin{figure}[t!]
\begin{center}
\includegraphics[width=.5 \textwidth]{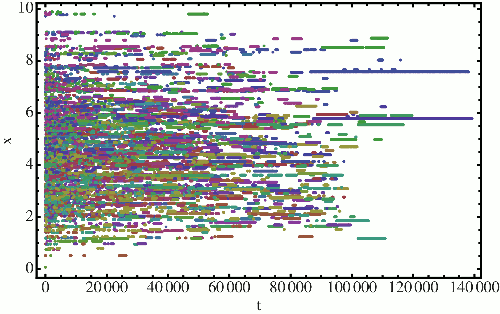}
\end{center}
\caption{Interaction of similar pasts between 40 past-sets during $10^5$ events. Same parameters as Fig.~\ref{fig1} but larger number of events. The longer run exhibits robust convergence to the lock-in phase and emergence of stable quasi-particles. 
}
\label{fig3}
\end{figure}

In Fig.~\ref{fig2} we increase the number of intervening pasts and in Fig.~\ref{fig3} the duration of interaction, to explore the depending of the degree of convergence. Results there confirm the hypothesis: given enough time models eventually evolve towards the lock-in and regular phase.

The length of time for emergence of regularity increases with increasing number of initial past-sets (which corresponds to larger initial variety). For constant number of past-sets, the degree of convergence increases with time, settling on the limit where only a few quasi-particles subsist interacting.

The general behaviour is thus an evolution from initial disordered behaviour, from which a few quasi particles eventually set off and interact only amongst themselves. The other pasts are left behind and stop interacting.  These results support the hypothesis that approximately time reversal symmetric behaviour can emerge from time asymmetric evolution rules.

\subsection{Details of the simulations}

In each run we initialize the system by picking the number of past-sets that will be interacting throughout evolution.  Each  of these past-sets is started off with an initial event which is given a random initial position at the common initial time, which is set to zero. Initially each past-set is empty. We pick $N$,  the total number of events that will be generated and pick an event generation rule regulating the interaction. 

We then generate $N$ events with the event generation rule.  The rule picks which two of the existing events are to be generated and which intersection of the null cones of these two events becomes the new event (because of the periodic identification of the spatial coordinates there are several possibilities.)  Thus, regardless of the rule there are always two events interacting and one event generated. 

As the number of pasts is fixed, each new event that is generated is placed in the past-set of one of the progenitors. We choose which progenitor will inherit the new generated event in its past: in our simulations we chose for simplicity this to be the progenitor which provided the left incoming photon to the new event. This causes no asymmetry in the results, except in Fig.~\ref{fig4}  where the effect is easily identifiable and has no significance for the results. 

So  each event belongs to one and one past-set alone. We could instead store the new event in both pasts, however this makes the algorithm computationally heavier and would affect the speed of convergence. 


We present results here from two event generation rules.  The first  is that given Eq.~\ref{min}, which selects as progenitors events which have the most similar pasts.  For the second rule we picked the opposite case, which is a rule that selects as progenitors the pair of events which have the least similar pasts.

\subsection{From irregularity to regularity}

We observed the emergence of the regular, locked-in phase in all models simulated, and for all interaction rules we studied. How can we explain this?

The emergence of regularity always seems to be related to the loss of novelty in the system. In our case the novelty element is the random number of cycles added at each event generation. In the beginning all pasts intervene in similar amounts, any pair interacts only twice and interaction moves on the to next pair. The information in each past is minimal and novelty, in the form of the random number of cycles added, plays a more determining role in the generation of new events than the interaction rule. However, with each event generation the information in each past-set builds up, and begins to weigh  increasingly more in the selection of pairs. Correspondingly novelty plays progressively a minor role, until there are no new positions created and the number of possible space positions is constant. This is represented by the emergence of quasi-particles, with regular trajectories, which we observe in all four figures. This is the lock-in phase which is identifiable by the regularity of structures in the network.
We observed that, quite surprisingly, this lock-in regular phase exists for the each of the interaction rules and initial conditions we considered.

The loss of novelty is not absolute. We always keep introducing random cycles $n$ at each new event generation, but at some stage this stops being sufficient to destabilize regularity. So there's an interplay between the random input and the regularity of the rule we apply. The observed tendency is for the regularity to eventually overcome the amount of diversity. \textit{Irreversible dynamics seems to have the tendency to evolve towards predictable, reversible evolution}.

The regular phase is not inert and not without dynamics. 
The quasi-particles, even if seemingly independent, are a product of interaction between their respective pasts, each co-dependent on one another. These quasi-particles, apparently still and inert, are in fact continually interacting between themselves, mutually generating events in each other's trajectories. In this regular phase pasts are interacting in stable equilibrium, which consists of a repeating sequence of a small number of pairs generating each others positions. 

%

\subsection{The lock in phase and dynamical systems}

In order to understand the emergence of the reversible-lock in phase, we'll relate it to the emergence of limit cycles in discrete dynamical systems\cite{DDS}. 
Note that the emergence of the regular phase doesn't depend on the exact positions of the initial events, just the order of their positions relative to each other in the initial $t=0$ slice. This is because the model has conformal symmetry given by 
(\ref{metric2}). This way we can choose 
initial conditions so that the initial $x$ values are evenly spaced around the circle. We can choose  units so that $\epsilon =\frac{1}{2\pi}$, so that the periodicity in $x$ is $L = 1$.   Then, the $N_{pasts}$ initial events are at 
$t=0$ and $x_I=\frac{I}{N_{pasts}}$, where $I \in \{ 1 , 2 , \ldots , N_{pasts} \}$.  Then the initial values of $z^+$ are  $z^+_I=\frac{I}{N_{pasts}}$, while the initial
values of the $z^-$ are  $z^-_I=-\frac{I}{N_{pasts}}$.

Under the evolution rules (\ref{interaction}) 
\begin{equation}
(z_{\rm new}^+,z_{\rm new}^-)=(z_{(1,2)}^++n\,\,z_{(2,1)}^--m )\,, 
\label{zmixup2}
\end{equation}
This means that the possible $x$ values can only come from the set 
\f
x \in \{ \frac{I}{N}   \}
\ff
for $I \in \{ 1 , 2 , \ldots , N_{pasts} \}$, which are the set of initial values.  
 Thus, the coordinates of the events can only come from a finite set of numbers.  Meanwhile, the momenta are just  traded around.  

Thus, the model can be mapped to a discrete dynamical system with a finite number of possible states.  
Such systems, it is well known, evolve to limit cycles\cite{DDS}.  These cycles, once entered, are locked in by the deterministic evolution rule that gives each state a unique successor.  As we will discuss in detail elsewhere, the capture of an irreversible dynamical system by a limit cycle can be seen to be a model of a transition from a generally irreversible dynamics to a reversible phase.  The general system is irreversible because each state may not have a unique predecessor, even if determinism requires each state to have a unique successor.  But once a limit cycle is entered, each state has both a unique predecessor and a unique successor.  Hence, restricted to the limit cycles, the system may be said to be governed by an effective reversible dynamics.  This will be discussed in more detail in \cite{DDS-ECS}.  


\subsection{Alternative interaction rule: Non locality}\label{max}

We might suspect that the reason we observe regularity emerging in the former scenario is because the interaction rule between pasts is local - it selects the two pasts which are most similar (which in our case means closest space-time histories). If this were the case the emergence of regularity and reversible evolution would simply reflect the local nature of the rule applied and have no particular meaning beyond that. For this reason we also 
investigated whether regularity emerges when a non local interaction rule is used. One example of a non local rule is the selection of the pasts which are maximally distinct, so we pick as progenitors pairs which maximize rather than minimize Eq.~(\ref{min}).

\begin{figure} [t!]
\centering
$\begin{array}{cc}
\includegraphics[width= 0.5\textwidth]{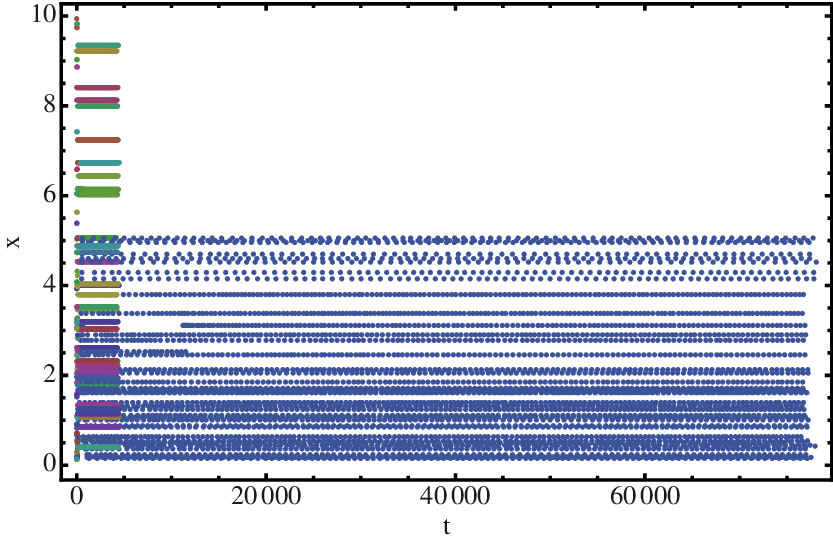}&
\includegraphics[width= 0.5\textwidth]{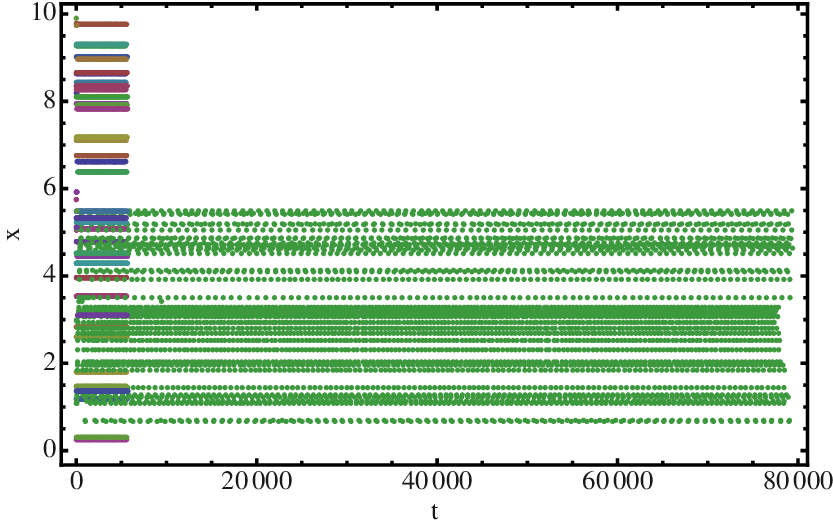}\\
\end{array}$
\caption{Non local interaction of maximally distinct pasts. Same parameters as Fig~\ref{fig1}, 40 past-sets interacting for $10^4$ events but interaction rule given by maximizing as opposed to minimizing Eq.~(\ref{min}). The asymmetry of the pattern is due to the asymmetry of our storage rule and of no further consequence. The same two phases of irregularity and lock-in occur but the emergence of quasi-particles, and consumption of diversity, is prompt in this non-local rule.}
\label{fig4} 
\end{figure}

In Fig.~\ref{fig4}  we plot the model generated under this rule\footnote{The observed asymmetry in space positions in Fig.~\ref{fig4}  reflects the asymmetry in our choice for the progenitor which inherits the new event - the one providing the left incoming photon - and has no further significance. If we had chosen to store events in both progenitors the entire plot would be filled with quasi-particles.}. 
Here we equally observe an initial period of irregular mixing where evolutions is akin to that in the minimum distance case.
Then a sudden lock-in takes place, characterized by the emergence of events with a single colour and hence past-set.  The emergence of regularity occurs promptly as opposed to progressively, and much sooner than with the minimally distinct rule. This is again a symptom that novelty was exhausted. We can explain this by noting that over time the expression for the past given by Eq.~(\ref{past}) becomes essentially dominated by the age of each past, given that time is always evolving and space coordinates are cyclic in $L$. 

Once age starts to dominate the past, and given that we are using the rule where maximally distinct pasts interact, current events - which are the oldest - begin to interact with the events that were left behind at the beginning - the youngest - creating ever bigger distances between any pair of events.  

The oldest past starts interacting only with the youngest pasts that had been left behind at the beginning, and stop interacting with other pasts of similar age. This transition is clearly visible in Fig.~\ref{fig4}. We see that initially, in the mixed colors stage, all pasts are interacting in similar proportion. When the lock-in occurs only one color is left, meaning that a single past - the oldest - starts to interact with all the youngest pasts. The trajectories of the emergent quasi particles correspond to those of the youngest pasts, that had not yet interacted. 

When lock-in occurs, the number of positions in the game 
becomes constant, no new positions are created. We end up with approximately as many quasi-particles as there are young unused pasts: those which were left behind at the beginning. No new pairs are created, novelty was exhausted and regularity emerged. 

We conclude that, contrary to what we might expect - that the emergence of regularity in the version where similar pasts interact is owed to the fact that the interaction rule is local - here the interaction rule is non local, and we still observe the emergence of regularity in the same fashion, and even reached more promptly than in the local model.

\subsection{Evolution in Cyclic or Extended Space}
\label{fold_unfold}

In the expression for generation of new event coordinates Eq.~(\ref{interaction}) the extra cycles $n\,\rm L$ must be included. If $n=0$ the number of different events generated in the future is limited -- there's only a finite combination of progenitor events $(z^+_I,z^-_I)$ and $(z^+_J,z^-_J)$. $N_{past}$ initial different events can generate maximum $N_{past}^2$ different new events. 
If cycles are not considered there is no possibility of photons interacting in future cycles other than the first, meaning they interact every time they cross. Since we are in $1+1$ dimensions this results in a simple system of oscillating photons, moving back and forth between each other.
This way in the absence of multiple cycles no new information (novelty) is inputed in the form of random number of cycles, event diversity is quickly exhausted and repeated, only a very small number of different events is generated.

With the addition of extra cycles a coordinate $x$ is always identified with $x+ n\rm L$.  
The addition of extra cycles allows us to investigate a different issue.
We could suspect that regularity in this compact space emerges because we are in confined space, in the sense that there is not enough diversity in spatial positions to create irregularity, and novelty is always used up.

\begin{figure} [t!]
\centering
$\begin{array}{cc}
\includegraphics[width= 0.5\textwidth]{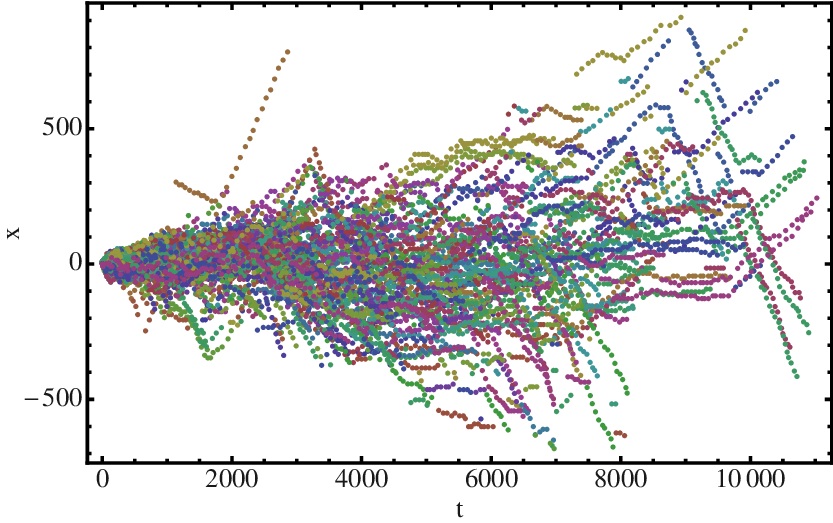}&
\includegraphics[width= 0.5\textwidth]{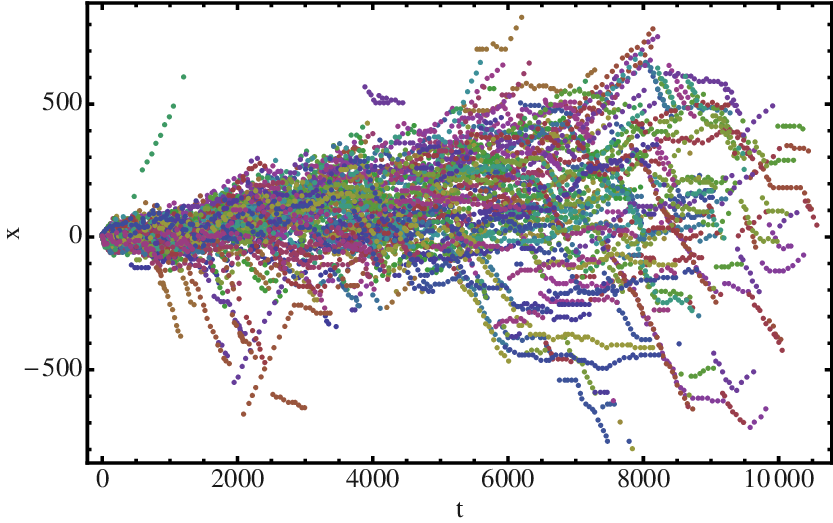}\\
\end{array}$
\caption{Interaction of similar pasts in extended space. 40 past-sets interacting for $10^4$ events. Same runs as Fig.~\ref{fig1} here represented in extended space where cyclic coordinates are not identified with each other. The quasi-particles are the diagonal lines. This representation in extended space also allows us to see that each quasi-particle is a product of interaction with another particle, and not inert.}
\label{fig5} 
\end{figure}

To counter check this we have examined interaction in an extended space, in which coordinates in different cycles are not identified, so the distance between events increases with each cycle. Fig.~\ref{fig5} shows this interaction. There we also observe that regularity, and the lock-in phase occur after enough events have taken place.
Quasi-particles emerge also in this interaction in extended space, represented by the diagonally evolving lines.

This shows that the emergence of the regular phase is independent of interaction in compact or extended space, and is not dependent on higher density of events in compact space nor on the sparser density in extended coordinates. 

Representing interaction in this extended space further allows us to confirm that quasi-particles are not static once they emerge, as we argued above, they are the result of two pasts interacting and mutually generating events in each other's trajectories. So the regular phase is in reality not inert and static.

\section{Summary and conclusions}

We began this work as we began the paper, by a search for principles that could guide the search for a physics that could be applied to the whole universe.
This must\footnote{As it is argued in detail in \cite{TR,SU}} be a novel form of dynamics that can be applied to a unique system-the universe as a whole, that is neither quantum mechanics nor general relativity, from which quantum physics and space-time emerge as approximate descriptions of systems that may be regarded to a sufficient degree of approximation as isolated subsystems.   

\subsection{Summary of ideas}

We formulated four principles, which we presented as {\bf Principles A to D} in section 1.  The first two have to do with the nature of time and assert that time, taken as the agency that produces new events from present events,  is truly fundamental and irreversible.  This focuses attention on an agent which  creates novel events-what we call the event generator,  The third principle asserts the philosophy of relationalism-  that space-time emerges from the action of that agent as a reflection of the network of causal relations that agent creates.  The fourth principle limits the reach of relationalism to assert that in addition to their relational, spatial-temporal properties, events are endowed by intrinsic properties, and these include energy and momenta.   

From these principles we draw several conclusions.
\begin{enumerate}
\item We work within an ontology of events, and the causal processes that create new events out  of old events.  

\item Each event should be uniquely distinct from the rest.   This is a consequence of Leibniz's principle of the identity of the indiscernible applied to an ontology of events and causal processes. 

\item The fundamental dynamics resides in the choices made by the events generator: which pairs (or small set of) present events will 
give rise to a new event and how are the properties of those progenitor events-both relational and intrinsic-transfered to the newly created event.\\
We should emphasize that no existing paradigm addresses the question of why particular events are created.  The existing causal set dynamics are either stochastic or based on a sum over histories approach to quantum physics.  Another paradigm based on an events ontology-the Feynman diagram approach to $QFT$-is also based on a sum over histories.  

\item The event generator must make use of and create the data that uniquely identifies each event.   
\end{enumerate}

This emphasis on the uniqueness of events, and its role in the fundamental dynamics, stands in strong contrast to the existing paradigms for dynamics, which assume that microscopic systems are simple and come in vast ensembles of identical copies.  This puts the emphasis on symmetry principles applied to identical elementary particles.   In contrast to this we assert that at the level of the elementary events and the dynamics that generates them every event is unique so there are no ensembles of identical events and \textit{no fundamental symmetries}.  

Consequently we assert that the kind of physics we have developed to this point, based on identical properties classified by symmetries, must emerge at an intermediate scale.  We can say more specifically that this emergence has to do with truncating the description within which each event is unique, so as to give an effective description of subsystems of the universe with minimal information.  

It is also natural to conjecture that this truncation is responsible for the statistical nature of quantum physics.  The stochastic description arises by neglecting data that renders each event unique which results in a description of emergent quasi-particles that fall into large classes of nearly identical copies.  It is at this emergent level that symmetry and identity arise as aspects of a statistical and approximate description.  We can express this by saying that there are hidden variables which reside in the data that makes each event unique and determines which events take place.  

\textit{Time reversal invariance is amongst the symmetries that must emerge from the coarse graining that neglects the unique identity of each event}.  From our perspective fundamental physics is time asymmetric and the apparent time symmetry of the laws of nature is approximate and emergent.  

We explored the implications of this new approach to fundamental physics by inventing a new kind of model of space-time physics: an energetic causal set, which differs from the usual causal sets by having events endowed with intrinsic energy and momenta, which are transmitted by causal processes. 
We discovered that under natural assumptions-a small enough number of progenitors-spacetime naturally emerges as an approximate description based on a stationary phase approximation that appears in a statistical description of these systems.  This addresses a long standing difficulty of causal set models to generate emergent low dimensional spacetimes. 

We can offer a suggestion of why space-time emerges, which is that it resolves a problem faced by the event generator, which is how to uniquely label each event in a manner that requires a small amount of information and so is computationally efficient.  At first, each event is
distinguished by its causal past, but to specify these in a large universe to the point they render each event unique would take a vast amount of information.  The emergent peacetime coordinates summarizes this information in a small set of numbers.    

\subsection{Future prospects}

We then ran numerical experiments to study this class of models in $1+1$ dimensions. 

We saw explicitly in these simulations how a time irreversible unique events model could underlie an emergent dynamics of repeatable, phenomena.  As time evolves we observe quasi-particles emerging with stable  position even as the event generator continues to make them interact and generating events in each other's trajectories.

We thus have the following tentative picture, the testing and exploration of which will be the goal of future work:  The fundamental dynamics generates a network of unique, never repeatable, events. Composite processes, such as could underlie elementary particles, are to be identified by searching for small blocks of events, which display patterns that repeat in the network. A  particle can be said to occur whenever we observe its respective characteristic block structure, which is its signature. By following different occurrences of the same pattern in the network we can identify an  emergent particle dynamics. 

A key goal of future work will be to identify the limit where the science of subsystems breaks down and uniqueness of cosmological events must be taken into account.
Locally the structures in the universal network appear identical, and we achieve the limit of repeatable experiments by truncating the description to subsystems. There is however a scale where this frequentist science breaks down, subsystems are no longer similar, and we need to extend a subsystem to include its unique history in order to obtain a consistent cosmological dynamics. 

In the model of unique  events, the present is defined as the set of open events, instants that are in the process of completing. In the model we described here, these are those events or instants that still have unused photons. So the present is not a single instant, but a set of instants. This has a parallel in the discussion of whether the present moment is thin or thick: in this view the present is thick, made of those events that are in the process of realizing themselves.


One important question to study is the nature of the set of present events.  
In the model we studied in the last section, the present is continually generated from interaction of open events. Even in the case of stationary quasi-particles, that have static space-time trajectories, they are in fact are in fact interacting, their trajectories are created mutually by interacting with each other. There is no absolute stillness, the agency of time is continual. Both in the emergence of quasi-particles and in the irreversible phase, at each moment, one event or two events are transported into the past, stop being part of the present, and a new event is created, taking their place, and is brought into the present. 

The amount of information in the current live events set characterizes the potential for novelty and diversity of the future network. There are multiple possibilities for how to quantity the amount of latent novelty in a system. In our model, novelty arises with the random number of cycles introduced at each event generation. As argued above, when this novelty is exhausted lock-in occurs and the regular phase is reached. In the regular phase the diversity in the system is constant and limited. It will be an interesting to make the injection of novelty vary in time and observe how the systems reacts with regards to its reversible or irreversible dynamics.

Our results suggest that the irreversibility persists so long as diversity in the system is abundant. If there is a fixed quantity of  novelty, the tendency is for it to be used up in the generation of irreversibility. When novelty is exhausted, the amount of diversity (however we choose to quantify it) becomes limited, and the dynamics regularizes, becoming patterned and reversible. 

It is notable to observe that the lock-in and consequent emergence of reversible dynamics is robust and occurs whether the interaction rule is local or non-local i.e. whether the progenitors of now events are chosen to have past-sets which are maximally similar or maximally diverse.

A goal for future work will be to understand in more detail how fundamental irreversibility gives rise to standard reversible laws; what is the quantity that represents novelty; and how does the transition to the reversible regime occur.

\section*{Acknowledgements}

We thank
Niayesh Afshordi, Eugenio Bianchi, Astrid Eichorn, Laurent Freidel, Ted Jacobson, Lise Kay, Flavio Mercati, Markus Mueller, Rafael Sorkin, Rob Spekkens 
and Roberto Mangabeira Unger for discussions. This research was supported in part by Perimeter Institute for Theoretical Physics. Research at Perimeter Institute is supported by the Government of Canada through Industry Canada and by the Province of Ontario through the Ministry of Research and Innovation. This research was also partly supported by grants from NSERC, FQXi and the John Templeton Foundation.
M.C.\ was supported by the EU FP7 grant PIIF-GA-2011-300606 and FCT grant SFRH/BPD/79284/201(Portugal).  


\end{document}